\documentclass[authoryear,preprint]{elsarticle}

\usepackage{hyperref}

\usepackage{array}
\usepackage{color}
\begin{document}

\begin{frontmatter}

\title{Possibility of microscopic liquid water formation at landing sites on Mars and their observational potential}
\author[rvt]{B. P\'al}
\ead{bernadett.pal12@gmail.com}

\author[focal]{\'A. Kereszturi}

\address[rvt]{E\"otv\"os Lor\'and University, P\'azm\'any P\'eter s\'et\'any 1/A, 1117 Budapest, Hungary}

\address[focal]{Research Centre for Astronomy and Earth Sciences, 9400 Sopron, Csatkai u. 6-8., Hungary}

\begin{abstract}
Microscopic liquid brines, especially calcium-perchlorate could emerge by deliquescence on Mars during night time hours. Using climate model computations and orbital humidity observations, the ideal periods and their annual plus daily characteristics at various past, current and future landing sites were compared. Such results provide context for future analysis and targeting the related observations by the next missions for Mars. Based on the analysis, at most (but not all) past missions' landing sites, microscopic brine could emerge during night time for different durations. Analysing the conditions at ExoMars rover's primary landing site at Oxia Planum, the best annual period was found to be between $L_s$ 115 - 225, and in $Local\hspace{0.1cm} Time$ 2 - 5, after midnight. In an ideal case, 4 hours of continuous liquid phase can emerge there. Local conditions might cause values to differ from those estimated by the model. Thermal inertia could especially make such differences (low TI values favour fast cooling and $\textrm{H}_2\textrm{O}$ cold trapping at loose surfaces) and the concentration of calcium-perchlorate salt in the regolith also influences the process (it might occur preferentially at long-term exposed surfaces without recent loose dust coverage). These factors should be taken into account while targeting future liquid water observations on Mars.
\end{abstract}

\begin{keyword}
Mars\sep liquid water\sep deliquescence\sep ExoMars\sep mission planning
\end{keyword}

\end{frontmatter}

\section{Introduction}

The most probable form of liquid water on Mars today and during most of the Amazonian period are the microscopic liquid layers formed by deliquescence or by the interfacial forces at the ice-mineral interface. Confirmation of their existence and the characterization of their consequence could be understood by in-situ analysis. Although remote data provide information on the environmental conditions for the emergence of these liquids, and Earth based laboratory measurements provide information on the micro-physical processes, it is difficult to extrapolate the real microscopic events inside the Martian regolith under the given exact composition, particle size etc. Thus, in-situ analysis could provide the most relevant information here. As near future surface missions aim for low latitude sites, the microscopic liquid between ice and minerals cannot be analysed, while there is a better chance for the observation of deliquescence driven liquefaction. In this work we exploit the potential emergence and observational possibilities of deliquescence driven microscopic liquid water formation at future ExoMars candidate landing sites, in order to give predictions on the targeting of observations there. As the main contribution of this work to the field, we present numerical estimations of a range of parameters regarding which seasonal phase, and local time provide the ideal combinations for the emergence of liquid and possibility for these observations at the candidate landing sites. For this we present the properties of of the climate modelling we used, correlate the results with the threshold values for microscopic liquid formation of Mars-relevant salty solutions, analyse the possibility of the emergence of these liquids at past and future landing sites, and finally predict ideal conditions for the identification of these liquids by the ExoMars rover. \\

Our work and this article could help optimize the laboratory measurements on Earth - what kind of tests and simulations should be run in the future, where and in what season landers should be sent out to Mars for the highest chance of finding and analysing water in its liquid form. Our results could also prove to be useful in the planning and optimization of measurements at the ExoMars rover's landing site, including the earlier candidate sites. \\

\subsection{\textbf{Background information}}

This chapter provides an overview of the reasons and background conditions for the possible emergence of liquid water on Mars and the approach we used to predict their occurrence in this study. There is an ongoing debate regarding the potential existence of liquid water on the surface of Mars. Certain theoretical models predict that liquid water may be present \citep{clow1987} \citep{haberle2001} \citep{hecht2002}, while some surface features favour the ephemeral existence of a liquid phase \citep{brass1980} \citep{Mellon2001} \citep{knauth2002} \citep{motazedian2003} \citep{kossacki2004} \citep{kereszturi2009} \citep{szynkiewicz2009} \citep{mcewen2014}. Computations favour the emergence of microscopic liquid water, especially brines \citep{mohlmann2004} \citep{kossacki2008} \citep{martinez2013}. The presence of brines is suggested by the observations at the Phoenix landing site and other locations \citep{kossacki2004} \citep{chevrier2009} \citep{hecht2009} \citep{hecht2009det} \citep{renno2009}, which may act as a possible agent for the formation of some recent flow-like features. \\

Recently, using the meteorological observations from the Curiosity rover, it was demonstrated that the night time conditions are favourable for the emergence of microscopic, thin, liquid film on the surface of hygroscopic mineral grains on the Martian surface and in the shallow subsurface \citep{torres2015}. Taking into account the above mentioned possibilities, it is worth calculating the possible emergence of such microscopic liquids on Mars in order to better orient instrument development, and also target in-situ analysis there. If such microscopic liquids emerge, it happens under restricted physical and chemical conditions, at limited temporal and spatial occurrence, thus suggesting parameters for the targeted analysis is highly important.\\

Under the current usually cold and dry Martian conditions, the \textbf{deliquescence process} could provide microscopic liquid water without ice. Hygroscopic salts could adsorb water vapour from the atmosphere where deliquescence (\textcolor{black}{transition from solid to aqueous phase}), or efflorescence (\textcolor{black}{transition from aqueous to solid phase}) happen \citep{gough2011}. The threshold limits for these processes are the \textcolor{black}{eutonic} relative humidity (\textcolor{black}{$\textrm{RH}_{\textrm{eut}}$}, where deliquescence starts), the deliquescence relative humidity (DRH, where deliquescence is complete) or efflorescence relative humidity (where the thin liquid layer is lost and the material turns to a solid phase). Deliquescence phase transition occurs when the local relative humidity (RH) is equal to or higher than the DRH. \\

We worked with the \textcolor{black}{eutonic} temperature and the water activity of a solution at the \textcolor{black}{eutonic} T values, which are presented in Table \ref{tab:salts}. Data regarding calcium-perchlorate ($\textrm{Ca(ClO}_4)_2$) were acquired from the work of \citep{toner2014}, the magnesium-perchlorate ($\textrm{Mg(ClO}_4)_2$) from \citep{mohlmann2011} and the values for calcium-chloride ($\textrm{CaCl}_2$) from \citep{davila2010}. The characteristics of the salts and brines with the most probable occurrance among the candidate ones are described below. Magnesium-perchlorate ($\textrm{Mg(ClO}_4)_2$), identified by the Phoenix \textcolor{black}{lander}, is one of the most analysed salts \citep{hecht2009det}. The small spheroids observed on the lander's robotic arm appeared to merge during the mission, which some scientists argue could be a sign that the spherules were liquid water \citep{renno2009}. These could have been the liquid brine form of the salt mentioned above, but unfortunately detailed observations are not available. 

\section{Methods}

We used observations and model-based estimations together to see the ideal locations and periods for brine appearance at past and future Martian landing sites. From our model-based data, we created daily graphs at the chosen landing sites to see if there are ideal time periods for liquid water to occur. We used observational data as a starting point, at what $L_s$ values to start searching for these ideal periods, and also as reference data, to check if our model-based findings are in accordance with observations. We discuss these in detail below.\\

The model-based approach was provided by the \textbf{Mars Climate Database} (MCD) 5.1 version \citep{forget2011}. The MCD is a database of atmospheric statistics compiled from the Global Climate Model (GCM) simulations of the atmospheric circulation. The model takes radiative transfer through the gaseous atmosphere, dust and ice aerosols into consideration, while also simulating $\textrm{CO}_2$ ice condensation, the water cycle and numerous other important factors in representing the Martian atmosphere as accurately as currently possible. The model takes permanent ice reservoirs, the amount of permanent ice and seasonal frost into consideration, but the diffusion of water vapor into the regolith is not included \citep{navarro2014}. \\

We acquired data from the database by retrieving daily and yearly temperature (K), surface pressure (Pa)  and water vapour volume mixing ratio (mol/mol) values at past and future landing sites at 5 m above the surface. We used these to compute relative humidity with the following formula: 

\begin{equation}
Q_{sat} = \frac{100}{P} \cdot 10^b
\label{eq:qsat}
\end{equation}

$$ b = 2.07023-0.00320991 \cdot T - \frac{2484.896}{T} + 3.56654\cdot log(T) $$

where P is the \textcolor{black}{total atmospheric pressure at the surface} (Pa), and T is the temperature (K). Dividing water vapour volume mixing ratio by $Q_{sat}$  results in the relative humidity values  \citep{forget1999}. The $b$ parameter considers how the temperature affects the relative humidity values \citep{forget2015}. \textcolor{black}{It is important to note, that because of the large diurnal temperature swings, relative humidity and saturation can vary between a few percent even if the mixing ratio does not change at all. Therefore it is possible, that the calculation would underestimate the occurrence of high humidity conditions.} For the former landing sites we examined these conditions at $L_s$ (0,30,60...360), for the ExoMars future landing site, Oxia Planum at $L_s$ (0,5,10,15..360) and for the two backup landing sites, Aram Dorsum and Mawrth Vallis at $L_s$ (0,15,30,45...360). We collected data at 5 m above the surface, because this is the lowest altitude calculated. Below this height the data is extrapolated by the model, this contains certain levels of uncertainty. Strictly speaking, the value 1 m above the surface or less may not be completely accurate, if there is ice on the surface, and it could differ even if there is no ice present. The temperature of the surface is likely to be even colder during the night (\textcolor{black}{which is favourable for deliquescence to occur, if the temperature doesn't drop below the \textcolor{black}{eutonic} temperature}), and warmer during the day, than the 1 m values. Therefore our findings may not be absolutely accurate, but they most definitely give a good estimation on which periods and locations could be the most favourable for liquid $\textrm{H}_2\textrm{O}$ to appear, and other types of uncertainties, especially the behaviour of various salt systems, which could be occasionally larger. \\

Beside the modelling approach, we used the \textbf{Thermal Emission Spectrometer} (TES) dataset \citep{christensen1992} to analyse annual trends, with the "vanilla" software. This command line software, produced by the Arizona State University, reads and queries binary data from the TES dataset (from real observations), correlates between various data tables, and was used presently to acquire surface temperature values for daytime between 12-14 true solar time with a spatial resolution of approximately 3-8 km. As a result, these values can only be considered as a rough approximation of the \textcolor{black}{maximum diurnal} surface temperature. \\

During the TES data mining, we first created yearly average maximum and minimum temperature, pressure and relative humidity graphs and used these to determine which locations and seasons to further investigate. The relative humidity graphs were created by using water vapor data from \citep{smith2002}. To analyse the TES based values we used a simple, C-based program, which calculates average temperature/pressure/etc. depending on solar longitude values. Afterwards, another C-based program was used to smooth our curves by computing moving averages. The errors were estimated with the program also calculating standard deviations. These routines were used to generate our daily average temperature and relative humidity graphs at the chosen landing sites as well. \\

These TES based temperature, pressure and relative humidity data graphs were used as verification reference data, to check the validity of our MCD model-based findings. For the selected landing sites, we created daily temperature and relative humidity curves at every 30 LS from 0 to 360 (former landing sites), at every 15 LS (Aram Dorsum and Mawrth Vallis) and at every 5 LS for Oxia Planum, the future ExoMars landing site, with the model. In these daily graphs we searched for estimated temperature values higher than the examined salts \textcolor{black}{eutonic} temperature, as well as relative humidities above the DRH values. If both criteria were fulfilled, the given period is considered favourable for deliquescence. The fact that the DRH decreases at higher temperatures was not taken into account, therefore it is possible that deliquescence could occur more frequently, or for longer periods of time, than what we found. This search for favourable periods was also completed by a C++ program of our own. Based on our results, in this article we used $L_s$ 125 and 235 as representatives from our collected data from the MCD database. \\

We used relevant Mars \textbf{candidate salts and their brines} to estimate the possible emergence of the liquid phase. We examined numerous hygroscopic salts, and three different examples are presented below to see the values needed for analysis (Table \ref{tab:salts}). In our current study, we determined the most favourable periods by maximal daytime and minimal night time surface temperature, pressure and relative humidity values. In the future we plan on optimizing the MCD model we used, to compute more complex atmospheric circulation and various heating patterns of sunlit slopes more accurately. \\
\begin{table}[h!]
\centering
\small
\begin{tabular}{|c|c|c|} 
\hline
Salt & \textcolor{black}{eutonic} temperature & Water activity \\ \hline
$\textrm{Ca(ClO}_4)_2$ & 199 K & 0.51 \\ \hline
$\textrm{Mg(ClO}_4)_2$ & 212 K & 0.53 \\ \hline
$\textrm{CaCl}_2$ & 223 K & 0.62 \\ \hline
\end{tabular}
\caption{\label{tab:salts}Parameters of the examined salts, and water activity of the \textcolor{black}{eutonic} solution}
\end{table}
\section{Results}

\begin{figure}[h!]
\centering
	\includegraphics[width=10cm]{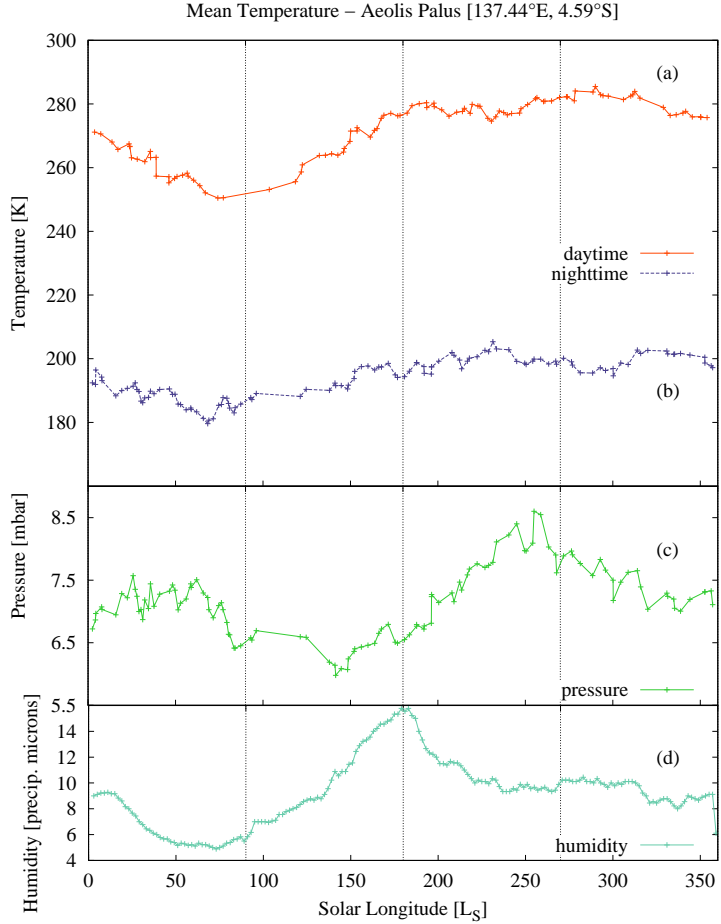}
	\caption{\label{fig:tesCU}\textcolor{black}{Annual mean daily maximum and minimum daytime (a) and night time (b) tem-
peratures, and annual change of atmospheric pressure (c) and humidity (d)} curves at the Curiosity rover's landing site. Data was extracted from the TES database. The annual averages were calculated by using our own C-based moving average program on the acquired measurement data. These measurement curves were used later as a verification reference to our model-based findings.}
\end{figure}

\begin{figure}[h!]
\centering
\includegraphics[width=10cm]{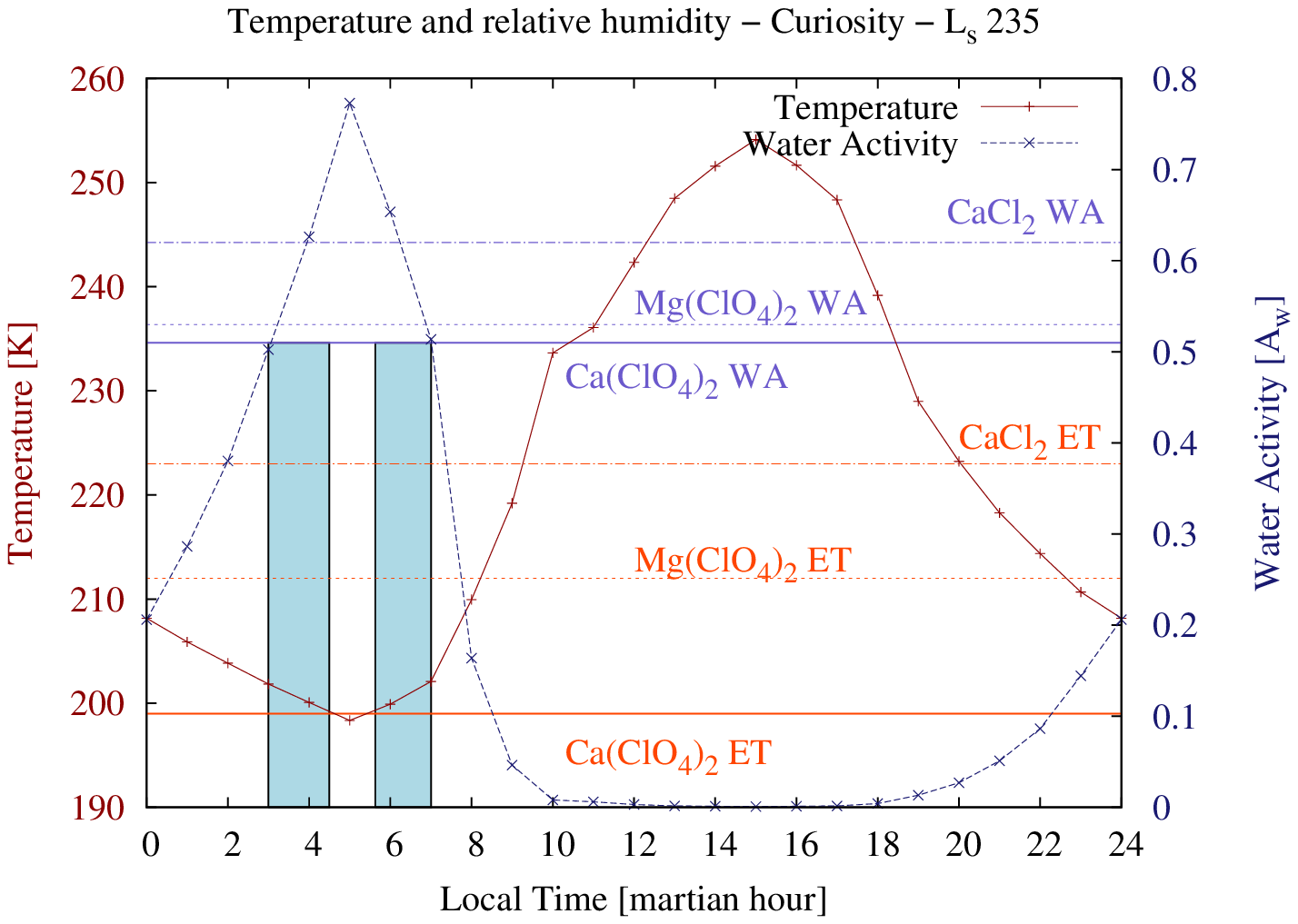}
\caption{\label{fig:cu235}The daily temperature (red) and relative humidity (blue) curves for the landing site of Curiosity. \textcolor{black}{The eutonic temperatures (ET) and water activity (WA) levels necessary for the salts illustrated are denoted by horizontal blue and orange lines. The lines have different matching line styles for each salt, with the salt's name on the line for identification.} There are two periods at the early and late part of the night (e.g. before and after the temperature minimum) when the conditions are favourable for brine formation for $\textrm{Ca(ClO}_4)_2$. These time periods are denoted by blue boxes. The left y axis shows the temperature in Kelvin, the right y axis the water activity values and the x axis the local time in martian hours.}
\end{figure}

To identify the possibility of deliquescence at the former landing sites, annual curves were computed first to see the most favourable seasonal periods (Figure \ref{fig:tesCU}). Having the ideal annual period selected, daily simulated curves were produced for the given location and seasonal phase - an example is Figure \ref{fig:cu235}. The simulated temperature and relative humidity values for the chosen landing site and solar longitude are indicated by red (temperature) and blue (relative humidity) lines. The \textcolor{black}{eutonic} temperatures of the particular salts are represented by orange lines with various styles for every salt. The water activity levels are similar to the \textcolor{black}{eutonic} temperature markers, but drawn with blue lines. The periods that seem to be favourable for deliquescence to occur are represented by light blue rectangles, marking the beginning and the end of the presumed ideal time period. The conditions are assumed to be ideal, if both the temperature and relative humidity values are above the salt's given \textcolor{black}{eutonic} temperature and water activity parameters. If these are true, then theoretically it is possible for liquid $\textrm{H}_2\textrm{O}$ to appear.  \\
Some example curves can be seen in Figure \ref{fig:tesCU} for the Gale crater (landing site of the Curiosity rover) to visualize the analyzed changes. Based on the comparison of different sites analyzed in this study, one should focus on the humidity at relatively low and middle latitude terrains, as this seems to be the most restricting factor, the ideal period here is around $L_s$ 180. \textcolor{black}{However the elevated atmospheric water does not necessarily mean that the relative humidity is also high. To find the ideal periods with elevated relative humidity values we used our model-based calculations.} \\

Figure \ref{fig:cu235} shows some daily profile examples of temperature and humidity for the Curiosity landing site at $L_s$ 235, together with the \textcolor{black}{eutonic} temperature and deliquescence humidity of $\textrm{CaCl}_2$, $\textrm{Mg(ClO}_4)_2$ and $\textrm{Ca(ClO}_4)_2$ systems. The daily trend has the following characteristics: daytime temperatures are too high to have an elevated RH value, but during night time it is low enough for these \textcolor{black}{eutonic} systems to have liquid phase. As well as many other locations, the ideal periods are in night time, in many cases at the following two periods: before and after midnight. Resemble trends will be seen later at different landing sites. For pure water, the conditions are not favourable: no possibility exists for its emergence, because the humidity is elevated when the temperature is far below the freezing point. \\

\subsection{\textbf{Considered previous landing sites}}

To see the possibilities, and conditions, and get a context for selecting conditions for the emergence of liquid water at future landing sites, we surveyed the previous ones. It is also interesting to compare them, in terms of which are the best and worst sites regarding this issue, and \textcolor{black}{when should the observations be carried out to potentially indentify whether deliquescence existed.} In Table \ref{tab:landingsites} former landing sites' parameters are indicated where we made analysis for the possible emergence of liquid. \\

\begin{table}[h!]
	\centering
	\small
	 \setlength\tabcolsep{1pt}
	\begin{tabular}{|c|c|c|c|c|}
	\hline
	\textbf{Lander} & \textbf{Longitude} & \textbf{Latitude} & \textbf{Landing date} & \textbf{End of mission} \\ \hline
	Mars 3 & $202^{\circ}$ E & $45^{\circ}$ S & 1971.12.02 & 1971.12.02 \\ \hline
	Mars 6 & $340^{\circ}$ 34' 48" E & $23^{\circ}$ 54' S & 1974.03.12 & 1974.03.12 \\ \hline
	Viking 1 & $311^{\circ}$ 52' 12" & $22^{\circ}$ 41' 49" N & 1976.07.20 & 1982.11.13 \\ \hline
	Viking 2 & $134^{\circ}$ 24' 36" & $48^{\circ}$ 16' 8" N & 1976.09.03 & 1980.04.11 \\ \hline
	Pathfinder & $326^{\circ}$ 52' 12" E & $19^{\circ}$ 7' 48" N & 1997.07.04 & 1997.09.27 \\ \hline
	Opportunity & $354^{\circ}$ 28' 12" E & $1^{\circ}$ 57' 0" S & 2004.01.25 & ongoing \\ \hline	 
	Spirit & $175^{\circ}$ 28' 43" E & $14^{\circ}$ 34' 18" S & 2004.01.03 & 2010.03.22 \\ \hline	
	Phoenix & $234^{\circ}$ 15' 3" E & $68^{\circ}$ 13' 8" N & 2008.05.25 & 2008.11.02 \\ \hline
	Curiosity & $137^{\circ}$ 26' 20" E & $4^{\circ}$ 35' 22" S & 2012.08.06 & ongoing \\ \hline
	\end{tabular}
	\caption{\label{tab:landingsites}List of former and ongoing Mars surface missions.}
\end{table}

In Table \ref{tab:favtime3} we present the favourable time periods according to our findings. In Table \ref{tab:favtime3} these favourable periods are those where the conditions are assumed to be ideal for the deliquescence of at least one of the three examined salts at the former Mars rover landing sites. \\

\begin{table}[h!]
	\centering
	\small
    \setlength\tabcolsep{0.7pt}
	\begin{tabular}{|c|c|c|c|c|c|c|c|c|c|c|c|c|} \hline
	\textbf{Lander} & \multicolumn{2}{c|}{Curiosity} & \multicolumn{2}{c|}{Opportunity} & \multicolumn{2}{c|}{Phoenix} & \multicolumn{2}{c|}{Spirit} & \multicolumn{2}{c|}{Viking 1} & \multicolumn{2}{c|}{Viking 2} \\ \hline
	\textbf{$L_s$} & 125 & 235 & 125 & 235 & 125 & 235 & 125 & 235 & 125 & 235 & 125 & 235 \\ \hline
	\textbf{$t1_{start}$} & 08:00 & 03:00 & 07:44 & 04:54 & 05:59 &  & 08:46 & & 03:40 & & 06:11 & \\ \hline	
	\textbf{$t1_{end}$} & 08:40 & 04:30 & 07:57 & 05:08 & 08:45 & & 08:57 & & 04:17 & & 07:48 & \\ \hline
	\textbf{$\Delta t$} & 0:40 & 1:30 & 0:13 & 0:14 & 2:46 & & 0:11 & & 0:37 & & 1:38 & \\ \hline
	\textbf{$T_{max}$} & 206 K & 202 K & 200 K & 200 K & 214 K & & 199 K & & 200 K & & 211 K & \\ \hline
	\textbf{$RH_{max}$} & 0.83 & 0.7 & 0.64 & 0.52 & \textcolor{black}{o.s.} & & 0.61 & & 1.54 & & 1.44 & \\ \hline
	\textbf{$t2_{start}$} & 21:00 & 05:38 & 23:06 & & 20:34 & & & & & & 21:36 & \\ \hline
	\textbf{$t2_{end}$} & 23:00 & 07:00 & 24:00 & & 23:10 & & & & & & 01:00 & \\ \hline
	\textbf{$\Delta t$} & 2:00 & 1:22 & 0:54 & & 2:36 & & & & & & 3:24 & \\ \hline
	\textbf{$T_{max}$} & 205 K & 202 K & 201 K & & 214 K & & & & & & 209 K & \\ \hline
	\textbf{$RH_{max}$} & 0.87 & 0.69 & 0.64 & & \textcolor{black}{o.s.} & & & & & & 0.96 & \\ \hline
	\end{tabular} 
	\caption{\label{tab:favtime3}List of ideal periods at former landing sites \textcolor{black}{according to our model-based findings}. The warmer and colder seasons are represented by $L_s$ 125 and $L_s$ 235 respectively. \textbf{$L_s$} stands for solar longitude, \textbf{$t1_{start}$} / \textbf{$t2_{start}$} and \textbf{$t1_{end}$} / \textbf{$t2_{end}$} are the beginning and the end of the time period (local solar time, where t1 is usually early morning and t2 for late night periods), \textbf{$\Delta t$} is the duration (hour:minutes, in Earth based values), \textbf{$T_{max}$} is the maximum temperature and \textbf{$RH_{max}$}  is the maximum relative humidity for the given time period. \textcolor{black}{The abbreviation o.s. means that the atmosphere is oversaturated.}}
\end{table}

It can be seen in Table \ref{tab:favtime3}, that certain chance existed for deliquescence at almost every former rover landing site. The widest occurrence along the seasons (at $L_s$ 125 or $L_s$ 235) of finding liquid water would be at the Curiosity rover's landing site, because both “seasons” we used for representative values proved to be favourable in the early morning hours, as well as the late night hours. It further improves our chances of finding liquid water that according to our predictions, these favourable time periods could even be two hours long. We also have good chances of finding water at the Opportunity landing site, but the ideal time periods are a bit shorter there, around 54 minutes maximum at night. At the Phoenix and Viking 2 landing sites there are ideal periods both in the morning and at night during the warmer season (represented by $L_s$ 125), and for a reasonably long period of more than 2 hours at a time. At the landing sites of Spirit and Viking 1 there is only a small chance of deliquescence to occur during the warmer season (represented by $L_s$ 125 for Viking 1, $L_s$ 235 for Spirit landing sites), in the morning or early morning hours, only for short periods at a time. \\

\begin{figure}[h!]
\centering
\includegraphics[width=10cm]{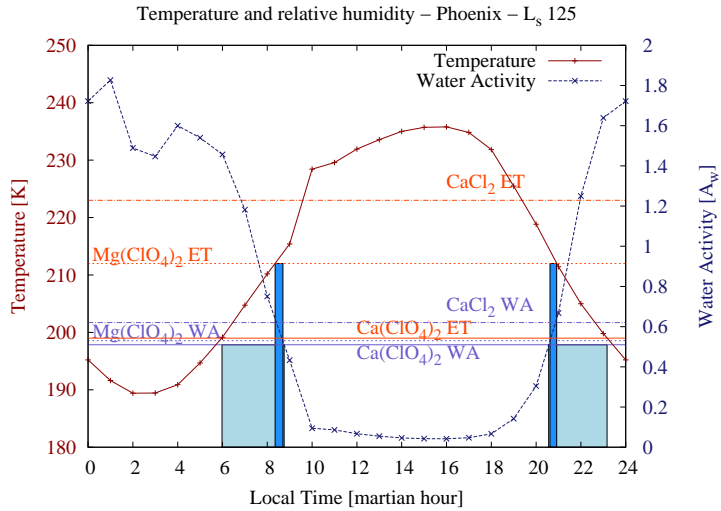}
\caption{\label{fig:PH125}Possible occurence of liquid $\textrm{CaCl}_2$, $\textrm{Mg(ClO}_4)_2$ and $\textrm{Ca(ClO}_4)_2$ brines at the landing site of Phoenix ($L_s$ 125). Daily temperature and humidity curves are marked with red and blue lines respectively. \textcolor{black}{The eutonic temperatures (ET) and water activity (WA) levels necessary for the salts illustrated are denoted by horizontal blue and orange lines. The lines have different matching line styles for each salt, with the salt's name on the line for identification.} The light blue boxes (different shades of blue for different salts) mark the night time periods when deliquescence could occur. The left y axis shows the temperature in Kelvin, the right y axis the water activity values and the x axis the local time in martian hours.}
\end{figure}

\begin{figure}[h!]
\centering
\includegraphics[width=10cm]{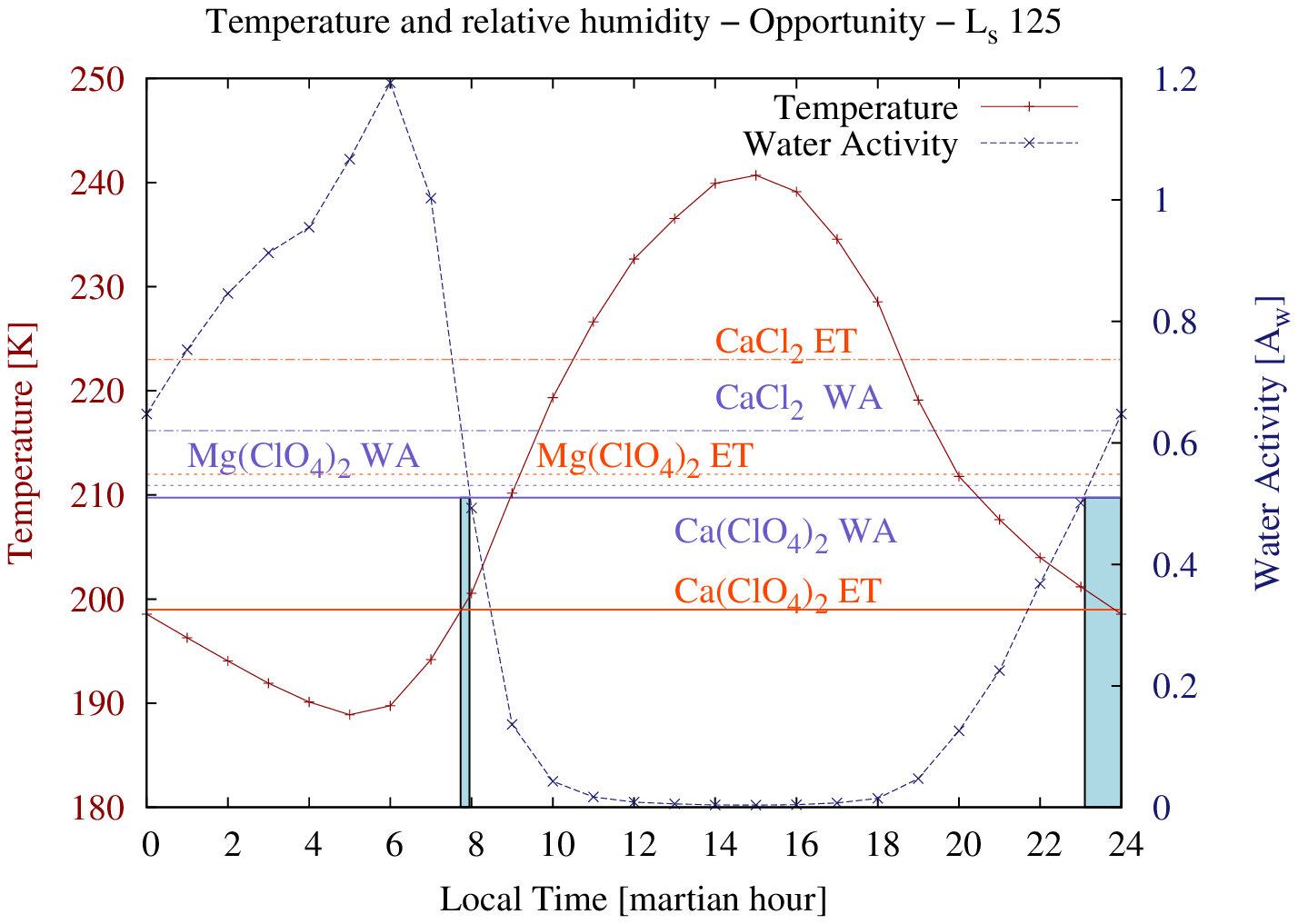}
\caption{\label{fig:OP125}Possible occurence of liquid $\textrm{CaCl}_2$, $\textrm{Mg(ClO}_4)_2$ and $\textrm{Ca(ClO}_4)_2$ brines at the landing site of Opportunity ($L_s$ 125). Daily temperature and humidity curves are marked with red and blue lines respectively. \textcolor{black}{The eutonic temperatures (ET) and water activity (WA) levels necessary for the salts illustrated are denoted by horizontal blue and orange lines. The lines have different matching line styles for each salt, with the salt's name on the line for identification.} The light blue boxes mark the night time periods when deliquescence could occur. The left y axis shows the temperature in Kelvin, the right y axis the water activity values and the x axis the local time in martian hours.}
\end{figure}

\begin{figure}[h!]
\centering
\includegraphics[width=10cm]{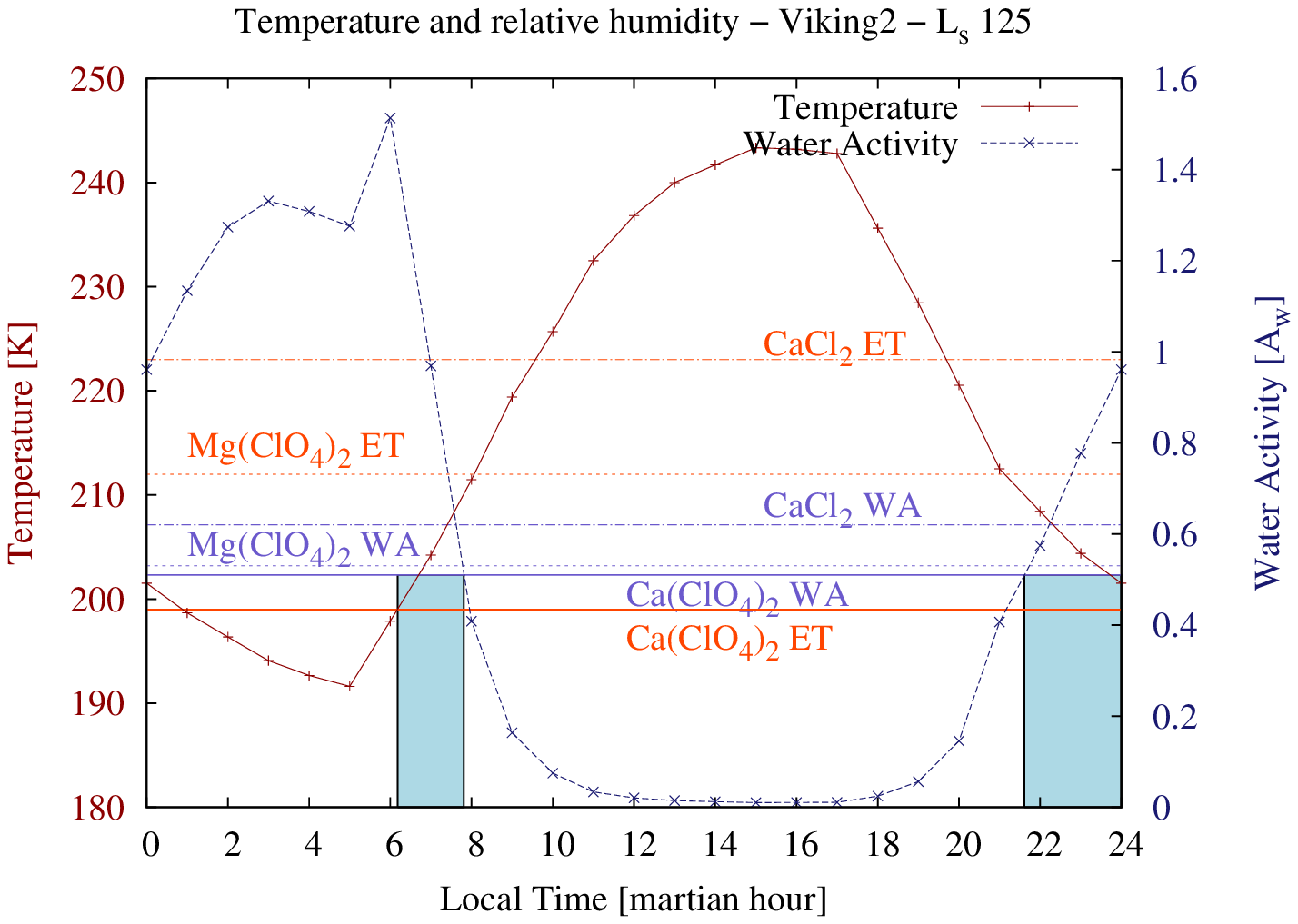}
\caption{\label{fig:V2125}Possible occurence of liquid $\textrm{CaCl}_2$, $\textrm{Mg(ClO}_4)_2$ and $\textrm{Ca(ClO}_4)_2$ brines at the landing site of Viking 2 ($L_s$ 125). Daily temperature and humidity curves are marked with red and blue lines respectively. \textcolor{black}{The eutonic temperatures (ET) and water activity (WA) levels necessary for the salts illustrated are denoted by horizontal blue and orange lines. The lines have different matching line styles for each salt, with the salt's name on the line for identification.} The light blue boxes mark the night time periods when deliquescence could occur. The left y axis shows the temperature in Kelvin, the right y axis the water activity values and the x axis the local time in martian hours.}
\end{figure}

\begin{figure}[h!]
\centering
\includegraphics[width=10cm]{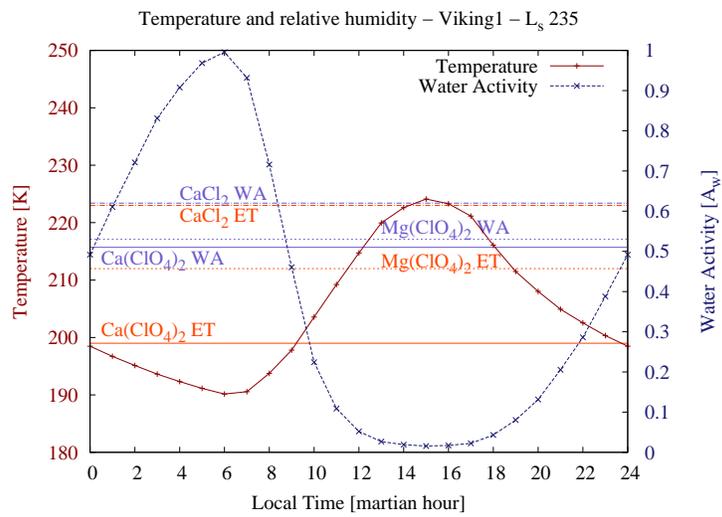}
\caption{\label{fig:V1235}Example of non-ideal circumstances at the Viking 1 landing site ($L_s$ 235) with respect to liquid $\textrm{CaCl}_2$, $\textrm{Mg(ClO}_4)_2$ and $\textrm{Ca(ClO}_4)_2$ brines to occur. Daily temperature and humidity curves are marked with red and blue lines respectively. The left y axis shows the temperature in Kelvin, the right y axis the water activity values and the x axis the local time in martian hours. \textcolor{black}{The eutonic temperatures (ET) and water activity (WA) levels necessary for the salts illustrated are denoted by horizontal blue and orange lines. The lines have different matching line styles for each salt, with the salt's name on the line for identification.} As we can see, there is no ideal period here according to our findings.}
\end{figure}

Some examples for the characteristics of the daily curves can be seen in Figure \ref{fig:PH125}, \ref{fig:OP125}, \ref{fig:V2125} and \ref{fig:V1235} with the following interesting observations: at the Phoenix landing site, both $\textrm{Ca(ClO}_4)_2$ and $\textrm{CaCl}_2$ brines could emerge, although the duration is longer and the chance is better for $\textrm{Ca(ClO}_4)_2$. Although the site of Viking 2 is located at a much lower latitude than Phoenix, almost the same duration of liquid appearance is possible at $L_s$ 125). \\

Based on the above mentioned findings, it is possible that during future Mars surface landings, microscopic liquid will emerge at certain periods, thus it is useful to prepare for possible observations. In the following overview, we discuss the observational possibilities at the ExoMars rover's landing sites, to see the ideal periods and evaluate influencing factors for possible targeted observations there. \\

\clearpage
\subsection{\textbf{Candidate ExoMars rover landing sites}}

To see the possibilities of liquid emergence for the ExoMars rover's future landing sites we performed the above analysis for the four candidate sites (Table \ref{tab:favtime4}). In Table \ref{tab:favtime4} we present the favourable time periods according to our findings, with the seasonal trends represented by $L_s$ 125 and 235. In Table \ref{tab:favtime4} the favourable periods are the times when the conditions are assumed to be ideal for the deliquescence of at least one of the three salts at the candidate ExoMars landing sites. \\

\begin{table}[h!]
	\centering
	\small
    \setlength\tabcolsep{2pt}
	\begin{tabular}{|c|c|c|c|c|c|c|c|c|} \hline
	\textbf{Location} & \multicolumn{2}{c|}{Aram Dorsum} & \multicolumn{2}{c|}{Hypanis Vallis} & \multicolumn{2}{c|}{Mawrth Vallis} & \multicolumn{2}{c|}{Oxia Planum}  \\
					 & \multicolumn{2}{c|}{7.9°N, 348.8°E} & \multicolumn{2}{c|}{11.9°N, 314°E} & \multicolumn{2}{c|}{22.3°N, 343.5°E} & \multicolumn{2}{c|}{18.20°N, 335.45°E} \\ \hline
	\textbf{$L_s$} & 125 & 235 & 125 & 235 & 125 & 235 & 125 & 235 \\ \hline
	\textbf{$t1_{start}$} & 07:26 & 03:21 & 02:40 &  & 06:24 &  & 02:44 & \\ \hline	
	\textbf{$t1_{end}$} & 07:32 & 04:13 & 02:47 &  & 07:06 & & 03:54 & \\ \hline
	\textbf{$\Delta t$} & 0:06 & 0:52 & 0:07 &  & 0:42 & & 0:10 & \\ \hline
	\textbf{$T_{max}$} & 200 K & 200 K & 199 K &  & 202 K & & 201 K & \\ \hline
	\textbf{$RH_{max}$} & 0.54 & 0.6 & 0.49 & & 0.67 & & 0.61 & \\ \hline
	\textbf{$t2_{start}$} & 23:31 & 06:00 & & & 20:28 & & &  \\ \hline
	\textbf{$t2_{end}$} & 24:00 & 07:00 & & & 01:52 & & & \\ \hline
	\textbf{$\Delta t$} & 0:29 & 1:00 & & & 2:25 & & & \\ \hline
	\textbf{$T_{max}$} & 202 K & 200 K & & & 204 K & & & \\ \hline
	\textbf{$RH_{max}$} & 0.57 & 0.6 & & & 0.76 & & & \\ \hline
	\end{tabular} 
	\caption{\label{tab:favtime4}List of ideal periods at candidate landing sites. The warmer and colder seasons are represented by $L_s$ 125 and $L_s$ 235. $L_s$ stands for solar longitude, \textbf{$t1_{start}$} / \textbf{$t2_{start}$} and \textbf{$t1_{end}$} / \textbf{$t2_{end}$} are the beginning and the end of the time period, \textbf{$\Delta t$} is the duration (hour:minutes), \textbf{$T_{max}$} is the maximum temperature and \textbf{$RH_{max}$} is the maximum relative humidity for the given time period.}
\end{table}

As it can be seen in Table \ref{tab:favtime4}, the favourable periods for deliquescence to most likely occur are in the early morning and late night hours. Out of the ExoMars rover candidate landing sites, Aram Dorsum, located at $7.9^{\circ}N$, $348.8^{\circ}E$ looks the most promising, with possible liquid stability periods available both during the night and early morning hours. At Mawrth Vallis, located at $22.3^{\circ}N$, $343.5^{\circ}E$ there's still a good chance for liquid water to appear in the morning hours, and at night, but only during the martian “summer season” (represented by $L_s$ 125). It is important to note though, that among the four candidate landing sites, the longest possible period for deliquescence to occur was found here. At the remaining two sites, Hypanis Vallis ($11.9^{\circ}N$, $314^{\circ}E$) and Oxia Planum ($18.20^{\circ}N$, $335.45^{\circ}E$) the ideal periods are relatively short, and appear only at night, during the warmer season. \\

Following the ESA decision, among the candidates, the Oxia Planum was selected as the primary landing site. Below we summarize the observational possibilities at that location to see the conditions for microscopic liquid appearance there, focusing firstly on the annual, secondly on the daily changes. \\

\begin{figure}[h!]
\centering
	\includegraphics[width=10cm]{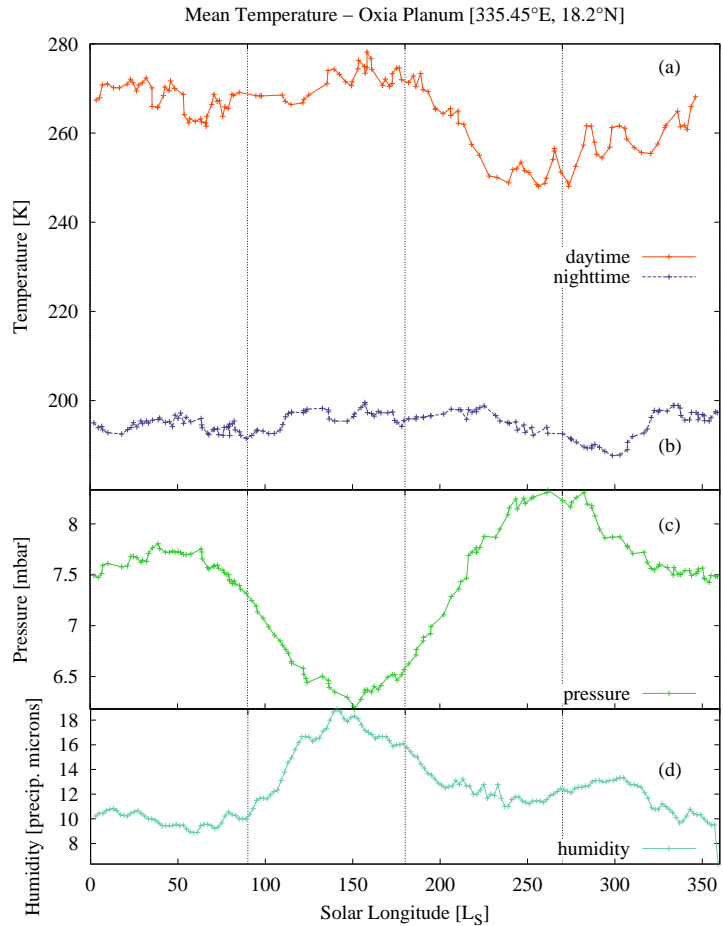}
	\caption{\label{fig:tesOX}Annual mean daily maximum and minimum daytime (a) and night time (b) temperatures, and annual change of atmospheric pressure (c) and humidity (d) for the primary landing site of the ExoMars rover.}
\end{figure}
Yearly mean daily maximum and minimum temperature, pressure and humidity values at Oxia Planum can be seen in Figure \ref{fig:tesOX}, extracted from the TES database. The annual averages were calculated by using our own C-based moving average program on the acquired measurement data. The annual curve shows that the night time temperatures do not change much through the year, while the change in the humidity shows typical annual trends. The ideal period in this aspect could be roughly between $L_s$ 100 - 225. During this period the temperatures during the day (6 - 18 local time) vary between 198 K and 244 K on average, and during the night between 198 K and 227 K on average. \textcolor{black}{However the elevated atmospheric water does not necessarily mean that the relative humidity level is also high. To find the periods with high relative humidity we used our model-based calculations seen below.} \\

\begin{figure}[h!]
\centering
\includegraphics[width=10cm]{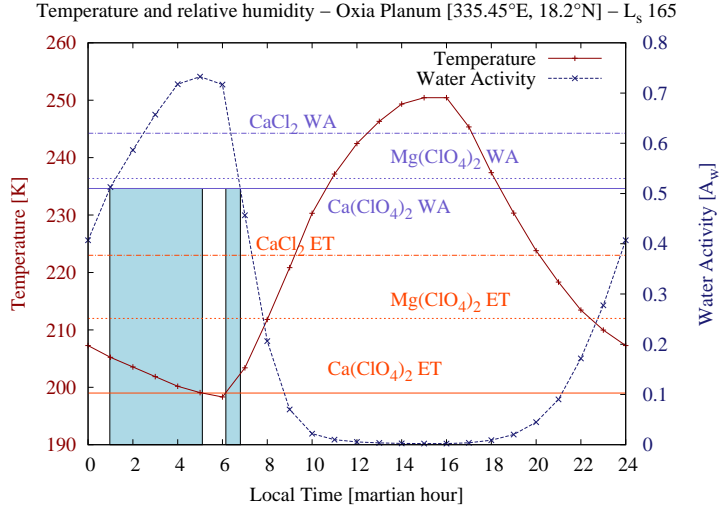}
\caption{\label{fig:ox165}Possible occurence of liquid $\textrm{CaCl}_2$, $\textrm{Mg(ClO}_4)_2$ and $\textrm{Ca(ClO}_4)_2$ brines at Oxia Planum at $L_s$ 165. Daily temperature and humidity curves are marked with red and blue lines respectively. \textcolor{black}{The eutonic temperatures (ET) and water activity (WA) levels necessary for the salts illustrated are denoted by horizontal blue and orange lines. The lines have different matching line styles for each salt, with the salt's name on the line for identification.} The light blue boxes mark the night time periods when deliquescence could occur. The left y axis shows the temperature in Kelvin, the right y axis the water activity values and the x axis the local time in martian hours.}
\end{figure}

Typical daily curves of temperature and humidity for the day with optimal conditions for deliquescence to occur at Oxia Planum  is at $L_s$ 165 - this can be seen in Figure \ref{fig:ox165}. It shows that the temperature is above the \textcolor{black}{eutonic} value for $\textrm{Ca(ClO}_4)_2$, while the humidity is also above the threshold value for approximately 4 hours and 46 minutes in total between 1 am and 7 am. These ideal periods are marked with light blue rectangles. \\

\begin{figure}[h!]
\centering
	\includegraphics[width=10cm]{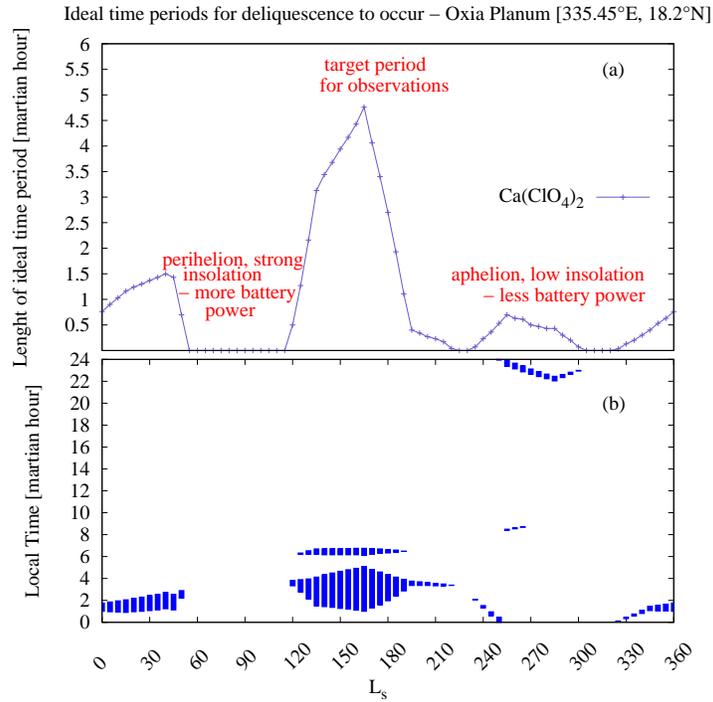}
	\caption{\label{fig:periodsOX}The time periods during which a liquid water phase may exist at the Oxia Planum landing (a inset), where the cumulative duration of deliquescence liquid phase can be seen. On the y axis the length of ideal time period is shown in martian hours, on the x axis, you can see the $L_s$ values from 0 to 360. In inset b daily periods after midnight are indicated at the bottom, and early evening periods after sunset at the top. On the y axis you can see the local time in martian hours, and on the x axis the $L_s$ values from 0 to 360.}
\end{figure}

The favourable periods of occurrence throughout the annual cycle are visualized in Figure \ref{fig:periodsOX} at the future ExoMars landing site. It can be seen that the best chance of finding liquid water is roughly from $L_s$ 105 to $L_s$ 225. The ideal periods are usually after midnight and in the early morning hours, between 1 am and 7 am. Between $L_s$ 255 - 300 there is also a possibility for deliquescence to occur before midnight in late evening hours, between 22 pm - 24 am. Texts also mark the seasons, indicating that the night time observation should be made after winter but still before summer (with moderate insolation at the solar panels). This is important in order to plan any related surface observations. \\
\pagebreak
\section{Discussion}
Based on the results from observations, and using model computations, it can be seen that a microscopic liquid film on the surface of hygroscopic salts might appear according to the following basic characteristics. Usually the early morning and late night hours are expected to be favourable for $\textrm{H}_2\textrm{O}$ to appear in liquid form. This liquid needs to be only microscopic scale in volume as there is only small vapour content in the atmosphere. During morning hours the ideal conditions last only for shorter periods of time, in the evening longer periods are more frequent - at least one but rather two or more hours long. \\

Regarding the specific landing sites, our calculation is in agreement with the observations at the Curiosity and Phoenix landing sites \citep{torres2015} \citep{renno2009} which were also compatible with the emergence of liquid water. Beyond these occurences, the chances were good at the Viking 2 landing site, where at $L_s$ 125 the temperatures were only 4-7 K lower, than the magnesium perchlorate's \textcolor{black}{eutonic} temperature and the relative humidity was higher than the salt's minimum water activity. This means, that the conditions could be fitting, if we consider such a small difference to be a possible error of the model used. The same time period could be ideal for the calcium perchlorate as well, the temperature difference is only 5 K. \\

Regarding the candidate salt systems, it is most likely for $\textrm{Ca(ClO}_4)_2$ to form liquid brines. The reason is mainly the salt's 199 K \textcolor{black}{eutonic} temperature, which is much lower, than any other's. The conditions could be favourable for $\textrm{Mg(ClO}_4)_2$, magnesium perchlorate to form liquid brines as well at the Phoenix rover's landing site, but only for a short amount of time. For the third examined salt, $\textrm{CaCl}_2$, calcium chloride none of the analysed periods were suitable, the temperatures remained 30 K below it's \textcolor{black}{eutonic} temperature on average, when the relative humidity values were high enough. \\

For calcium perchlorate there are multiple occasions, where and when the circumstances are close to the conditions needed for liquid brine stability. These are Spirit ($L_s$ 125), Aram Dorsum ($L_s$ 125 and 235), Hypanis Vallis ($L_s$ 125 and 235), Mawrth Vallis ($L_s$ 235) and Oxia Planum ($L_s$ 125 and 235) with a maximum of 7 K and a minimum of 1 K deviation from the necessary \textcolor{black}{eutonic} temperature, when the relative humidity was over the salt's minimum required water activity level. This confirms, that it is most likely for calcium perchlorate to form liquid brines on the surface of Mars and also suggests that the ExoMars rover might have potential to observe deliquescence of calcium-perchlorate at all of the three proposed landing sites. Among them, the best chance of finding liquid is at Aram Dorsum, however the longest favourable time period for deliquescence to occur is at the Mawrth Vallis landing site, in the evening hours. \textcolor{black}{At the selected prime landing site, Oxia Planum there is also reasonable chance of finding liquid water. According to our findings, the best period would be between $L_s$ 125 and 180, when liquid form could exist for 3-4 hours in the early morning.}  \\

\subsection{\textbf{Evaluation of other parameters}}

The computations in this paper are partly based on modelling and partly on observations with low spatial resolution. Currently this is the best approach available for the given parameters, where substantial uncertainty exists, which might be favourable or unfavourable for the emergence of liquid a phase. In reality several other parameters might also influence the conditions for the emergence of the liquid. Below we evaluate them briefly, keeping in mind that such evaluation could be no more than a rough estimate, but worth noting as it may point to ideal sites and conditions, where using the available facilities (equipments of the rover) could be identified more easily. Our findings also support identifying directions for future development. \\

\textbf{Wind strength} influences the stability of the once emerged thin liquid which could stay longer at wind shaded locations. Using the assumption that the cold night time air preferably “flows and accumulates” at local lows, the air there could be oversaturated in night time and thus decrease the evaporation rate, what is strongly relative humidity dependent. Using this argument, wind shaded depressions could favour the extended existence of thin microscopic liquid on the surface of hygroscopic salts there. \\

\textbf{Local mineralogical conditions} and possible compositional differences might have the strongest role in the emergence and maintenance of this liquid on the grains' surface. To provide a location with elevated salt abundance, three factors are relevant during the Amazonian. \textbf{1}. Salt forming processes (mainly evaporation) could have a role, but not in bulk water (macroscopic water) as it is expected rarely during the Amazonian, although there are models for the emergence of bulk liquid and related \textcolor{black}{eutonic} freezing there \citep{changela2010}. Much better chance exists considering formerly (Hesperian or Noachian) accumulated salts, which might still be there because of the weak surface changes on Mars in general. \textbf{2}. Water ice accumulation, possible micro-scale melting and drying could influence the spatial distribution of salts on Mars \citep{toner2015}. Low latitude glaciations \citep{head2003} or latitude dependent mantle formation \citep{kreslavsky2002} \citep{head2003} could bring water ice at middle and low latitude terrains, although near the equator, ice formation is only expected around the volcanoes (which are not among the candidate sites). Latitude dependent mantle formed at middle and high terrains, thus higher latitudes than ExoMars candidate sites. Despite these issues, it is worth considering the possible earlier deposited water ice and its potential role on salt concentration at low latitude locations (for example neutron spectroscopy detected bound $\textrm{H}_2\textrm{O}$ at the equatorial zone \citep{christensen2006} \citep{bandfield2007} \citep{feldman2011}). Water ice in physical contact with rocks could support the hydration of certain minerals, and later dehydration or \textcolor{black}{eutonic} freezing or even evaporation could leave behind salts and enhance their spatial concentration, like it is assumed for the landing site of Phoenix \citep{elsenousy2015}, \citep{toner2015rev}. \textbf{3}. Several models only make an account to a microscopic scale of $\textrm{H}_2\textrm{O}$ migration during the Amazonian period, when microscopic scale liquid formation, seep and sublimation play a role at a cm or mm scale. In this case, capillary forces might produce elevated occurrence of salts at certain spatial locations, or inside a given layer in a vertically heterogeneous duricurst \citep{cull2010}. \\

\textbf{Slope exposure} might influence the emergence or maintenance of microscopic liquid only by the insolation, which matters close to sunrise or sunset. Exposure might increase the temperature a bit, getting closer to the threshold limit, but also increases the evaporation rate. The consequence of slope exposure is poorly known thus more detailed analysis is required here. \\

\subsection{\textbf{Observing potential}}
Based on the above mentioned issues, to identify the existence of liquid water at the candidate landing sites, the following issues should be in focus. Temporal targeting of observations according to solar longitude and local time should focus on $L_s$ 115 - 225 and 2 - 5 am local time. \\

Methods of observation need to focus on previously poorly exploited periods of night time hours, occasionally under fast changing environmental conditions (not far from sunrise). This might require a new methodological approach both for the realization of observations (in darkness) and both from an engineering point of view, as the available energy, various security constrains (low temperature should cause difficulties for certain electronics) and software-based autonomy protocols (probes often used to be in a hibernated state during this period) are different for night time and daytime periods. \\

Tackling such an unusual challenge might require specific tools and protocols,  but probably not impossible to fulfil, using the engineering capacity and technology available today. Successful observations could improve meteorology models and provide a better understanding of local conditions. Testing the microscopic brine occurrence at different conditions during daily cycles today, could shed light on the condensation and possible brine occurrence under a different climate, when certain physical conditions could exist for a longer period during the day than currently, due to orbital tilt changes. \\

\section{Conclusions}
The Correlation of model- and observation-based annual and daily temperature plus humidity values points to ideal periods for the possible emergence of microscopic liquid brine on Mars. These ideal periods are found at night time, when in cold and vapour oversaturated conditions hygroscopic minerals could produce deliquescence and thus a thin liquid layer on the salts' surfaces. In most cases, because of the slower night time cooling and faster early morning warming, the longest periods of liquid occurrence is before the temperature minimum during the night. The ideal brine for this process among the Mars-relevant candidate ones is $\textrm{Ca(ClO}_4)_2$. Analysis of past landing sites for possible brine emergence allowed insight into the general observational possibility of such liquid - although more focused laboratory tests are required. At most sites (Opportunity, Curiosity, Phoenix, Viking 2) deliquescence could occur in theory (using modelled meteorological values and assuming the existence of hygroscopic salts there). At the landing site of Viking 1 and candidate landing site Hypanis Vallis minimal, at Spirit's landing site little to no deliquescence is expected. \\

Comparing different landing sites, the longest daily duration of liquid was around 6 hours and 15 minutes (at Phoenix lander's landing site), and the longest seasonal phase during which it could appear is 0 - 360 degree in $L_s$ at Curiosity rover's landing site. These findings are consistent with some earlier observations on the possible emergence of microscopic liquid water in the form of candidate droplets at the Phoenix probe \citep{renno2009}, and the ideal conditions identified at Curiosity's night time observations \citep{torres2015}. Analyzing the observational possibility at the planned primary landing site of ESA's ExoMars rover mission (Oxia Planum), for night time brine emergence $L_s$ 115 - 225 and Local Time 1 - 5 am are the ideal values. Around the middle of this seasonal phase, deliquescence is expected to happen for 4 hours continuously. It is also worth making observations beside this ideal range, as those measurements could shed light on the behaviour and possible emergence of microscopic liquid on Mars under different climatic conditions. \\

Calculating for the two backup landing sites of the ExoMars rover (ideal for a possible launch later than 2018) different seasonal periods are ideal for night time deliquescence (see the numerical details in Table \ref{tab:periods}). The longest seasonal period, while deliquescence is expected at night time is probably present at Aram Dorsum, where deliquescence is expected throughout most of the year, while at Oxia Planum and Mawrth Vallis a total duration of three times less is expected between southern spring and summer. The longest total continuous duration of deliquescence (4.7, 3.6 hours) is expected at the Oxia and Mawrth sites, while a much shorter duration (up to 2.2 hours) is expected at Aram Dorsum. \\

Of course, local conditions might differ substantially from the above mentioned general findings, thus these predictions are mere suggestions. Evaluating the possible local conditions, strong slope winds (that could enhance evaporation) are not expected at the Oxia Planum region. For the condensation, low temperature surfaces are favourable by catching the atmospheric $\textrm{H}_2\textrm{O}$. These are mainly low \textbf{thermal inertia} regions, i.e. loose dust covered surfaces, and it is also expected that such poorly consolidated, loose dust or sand covered areas did not spend much time on the Martian surface in a static state, and thus they need not spend a long time there for the possible concentration of hygroscopic salts to accumulate by an evaporation-sublimation process. Searching for the ideal site for deliquescence on Mars, especially at the ExoMars landing site, these factors should be considered. \\

\begin{table}[h!]
	\centering
	\small
	\setlength\tabcolsep{2pt}
 	\begin{tabular}{|c|c|c|}
	\hline
	Characteristics & Value & Landing site \\ \hline
	Longest ideal $L_s$ & & Aeolis Palus \\ 
	period overall & 0-360 [$L_s$] & (Curiosity) \\ \hline
	Longest ideal continuous & & Green Valley \\ 
	time period overall & 6.26 \textit{[martian hours]} & (Phoenix) \\ \hline
	 						   & 115-225 [$L_s$] \textit{(total: 110)} & Oxia Planum \\ 
	Longest ideal $L_s$ period & 105-360, 0-75 [$L_s$] \textit{(total: 330)} & Aram Dorsum \\
							   & 90-195 [$L_s$] \textit{(total: 105)} & Mawrth Vallis \\ \hline
	Longest ideal continuous & 4.76 \textit{[martian hours]} & Oxia Planum \\
	time period & 2.24 \textit{[martian hours]} & Aram Dorsum \\
	 			& 3.66 \textit{[martian hours]} & Mawrth Vallis \\ \hline
	\end{tabular}
	\caption{\label{tab:periods}In this table you can see the longest overall time and $L_s$ periods ideal for deliquescence to occur among all former landing sites, and the longest time and $L_s$ ideal periods for the ExoMars primary (Oxia) and two backup (Aram, Mawrth) landing sites.}
\end{table}

It is an important question how near of far our model prediction is from reality on Mars. The global climate model used was not able to handle regional or local differences and specific effects, but at the same time, this is one of the best available methods to estimate current Martian meteorological conditions, and there is a good chance that it shows real trends, and useful suggestions to target future measurements.

\section{Acknowledgment}
This work was supported by the COOP-NN-116927 and the COST TD1308 funds.

\section*{References}

\bibliography{marsref}

\end{document}